\documentclass{article}

\usepackage{amsmath}
\usepackage{url}
\usepackage{graphicx}

\begin{document}

\title{Delayed Gambler's Ruin}
\author{Tomohisa Imai and Toru Ohira\\
Graduate School of Mathematics, Nagoya University, Japan}
\date{}
\maketitle

\begin{abstract}
  We present here a new extended model of the gambler's ruin problem by incorporating delays in receiving of rewards and paying of penalties. When there is a difference between two delays, an exact analysis of the ruin probability is difficult. We derive an approximate scheme to find an effective shift in the initial assets of the gambler. Through comparison against computer simulations, this approximation is shown to work for small differences between the two delays.
\end{abstract}

\section{Introduction}

 The gambler's ruin is a classic problem in probability theories\cite{bailey}. A gambler who
has an initial asset takes on betting under certain probabilities of win and loss, until he is broken or reaches to a specific asset level. One of the simplest models  can be described as a restricted random walk with two absorbing boundaries.
The position of the walker indicates his asset and the walker can probabilistically gain or lose a unit of his asset at each time step, and moves accordingly until he reaches either boundary. This model has been studied extensively and various  extensions are made. Inclusion of non-moving time steps, and extensions to n-players gambler's ruin are such examples\cite{gut1,gut2,rocha1,rocha2}.

We will propose and investigate yet another extension of this problem in this paper, which we term as ``Delayed Gambler's Ruin''(DRG). The main feature of this model is that it includes delays in the gain or loss of a unit in the gambler's asset. This reflects that, in reality, payments and/or incomes often does not take place immediately at the time of corresponding events, such as a purchase with a credit card. Our proposed model, thus, moves according to past results of gambling.

Delays in dynamics have also been studied over past 50 years, often with feedback control systems\cite{glassmackey1988,stepan1989}. 
It has been shown that even a simple first order ordinary differential equation can show rather complex behaviors with delays.
They are called ``Delay Differential Equations''. Stochastic elements are further introduced onto both dynamical models by ``Stochastic Delay Differential Equations'' \cite{kuckler1,frankbeek2001} or probabilistic models by ``Delayed Random Walks''\cite{ohiramilton,ohirayamane}. 
The DRG we propose here can be thought of one example of these models incorporating stochasticity and delay.

We will first describe our formulation of the DRG. 
We investigate the mode with a particular focus on how delays affect the probability of ruin. 
It will be shown both by analysis and by computer simulations that an approximation using averaging can describe the behavior of the model for small delays.

\section{Delayed Gambler's Ruin}

 Let us start with a brief description of the Gambler's Ruin. A gambler attends a gamble with the initial asset of $x$ points. 
At each bet, he either wins one point or lose one point with a given probability. He ends his betting either when his assets become zero ("broken") or reaches his intended level $A$.
We now define some notations to analyze this problem.
\vspace{1em}

\begin{itemize}

\item
$p$: The probability of the gambler's winning a point. 
     (We also set $q = 1-p$ as the probability of losing.)

\item
$U_{t}$: Gambler's asset after  $t$ betting.

\item
$P_{A}(x)$: Probability that a gambler with the initial asset of $x$ points to become broken.

\item
$X_{t} = \pm 1$: The result of betting at  $t$.

\end{itemize}

By mapping this problem to restricted symmetric simple random walks with each step as 
$X_t$ and with absorbing boundaries at $0, A$,  following results are known.
\[
P_{A}(x)=
\begin{cases}
\frac{A-x}{A} & (p=q={1 \over 2})\\
\frac{(\frac{q}{p})^A - (\frac{q}{p})^x}{(\frac{q}{p})^A-1} & (p \not= q)
\end{cases}
\]

We will focus how this probability of ruin is affected by inclusion of delay. 
Two delays are introduced into the above Gambler's Ruin model: 
\begin{itemize}

\item
$\tau_p$: delay in payment of a point. 

\item
$\tau_r$: delay in receipt of a point.

\end{itemize}

At each bet, the paying or receiving of the point is deferred with the above delays. 
We also denote the probability of ruin in the Delayed Gambler's Ruin as $P_{A}^{\tau_{r},\tau_{p}}(x)$, which we will focus on the following  analysis. We also denote the stopping time, 
the time duration of the gambler's betting (i.e., the time between the beginning to the end of his betting, either by broken or by reaching $A$), as $T_{\tau_{r},\tau_{p}}$

It turns out the important parameter in analyzing this model is the difference between these two delays:
\begin{itemize}

\item
$\theta = \tau_r - \tau_p$: the difference of delay in payment and receipt.

\end{itemize}

With the above setup, we now start our analysis by considering different cases of $\theta$.
In the following, we assume that the initial asset is further away from the boundary than the difference between delays.
\[
A-x>\mid \theta \mid\quad and \quad x>\mid \theta \mid
\]

\subsection{$\theta=0 \quad (\tau_{r}=\tau_{p})$}

We first consider the case of $\theta=0$, which means two delays are the same, $\tau_{r}=\tau_{p}$.
In this case, the gambler's asset $U_{t}$ just after the $t$-th betting is given as follows.  

\[
U_{t} =
\begin{cases}
x \quad (1\leq t \leq \tau_{r})\\
x+ \displaystyle \sum_{k=1}^{t-\tau_{r}}X_{k} \quad (\tau_{r} < t)
\end{cases}
\]

(We can replace $\tau_{r}$ by $\tau_{p}$.)

Naturally, the gambler's betting does not end at least before $t = \tau_{r}$. 
Hence, the stopping time satisfies $T_{\tau_{r},\tau_{p}} > \tau_{r}$, and the probability for him to be broken is given as

\[
P_{A}^{\tau_{r},\tau_{p}}(x) 
= P(U_{T_{\tau_{r},\tau_{p}}}=0)\\
= P\left( x+ \displaystyle \sum_{k=1}^{T_{\tau_{r},\tau_{p}}-\tau_{r}} X_{k} =0 \right)
\]

Let $T_{0,0}$ be the stopping time when there are no delays both in receipt and payment. 
Then, we can see from the definition of the model that

\[
T_{0,0} = T_{\tau_{r},\tau_{p}} - \tau_{r}.
\]

This means that we can reduce the problem for the case of $\theta=0$ to the original Gambler's Ruin, leading to the following ruin probability for $\tau_{r}= \tau_{p}$

\begin{eqnarray}
P_{A}^{\tau_{r},\tau_{p}}(x) &=& P(U_{T_{\tau_{r},\tau_{p}}}=0)\nonumber\\
&=& P\left( x+ \displaystyle \sum_{k=1}^{T_{\tau_{r},\tau_{p}}-\tau_{r}} X_{k} =0 \right)\nonumber\\
&=& P\left( x+ \displaystyle \sum_{k=1}^{T_{0,0}} X_{k} =0 \right)\nonumber\\
&=& 
\begin{cases}
\frac{A-x}{A} & (p=q={1 \over 2})\\
\frac{(\frac{q}{p})^A - (\frac{q}{p})^x}{(\frac{q}{p})^A-1} & (p \not= q)
\end{cases}
\label{sdelay}
\end{eqnarray}
\\

\subsection{ $\theta > 0 \quad (\tau_{r} > \tau_{p})$}

Let us now consider the case when the delay in receipt is longer than that of payment,
$\tau_{r} > \tau_{p}$ ($\theta>0$).

In this case the asset $U_{t}$ of the gambler at $t$ is given as follows.
\[
U_{t}=
\begin{cases}
x & (0\leq t \leq \tau_{p})\\
x-\displaystyle \sum_{k=1}^{t-\tau_{p}} \delta_{-1,X_{k}}  & (\tau_{p} < t \leq \tau_{r})\\
x+\left(\displaystyle \sum_{k=1}^{t-\tau_{r}} X_{k} \right)-\left( \displaystyle \sum_{k=t-\tau_{r}+1}^{t-\tau_{p}} \delta_{-1,X_{k}}  \right) & (\tau_{r} < t)
\end{cases}
\]

Here $\delta_{i,j}$ is a Kronecker delta,
\[
\delta_{i,j}=
\begin{cases}
1 & (i=j)\\
0 & (i \neq j)
\end{cases}
\]

Since we have set the initial condition as $x>\mid \theta \mid$, the stopping time $T_{\tau_{r},\tau_{p}}$ satisfies $T_{\tau_{r},\tau_{p}} > \tau_{r}$.
Hence, the probability of the ruin is formally written down as
\begin{eqnarray*}
P_{A}^{\tau_{r},\tau_{p}}(x) &=& P(U_{T_{\tau_{r},\tau_{p}}}=0)\\
&=&  P\left( x+\left( \displaystyle \sum_{k=1}^{T_{\tau_{r},\tau_{p}}-\tau_{r}} X_{k} \right)-\left( \displaystyle \sum_{k=T_{\tau_{r},\tau_{p}}-\tau_{r}+1}^{T_{\tau_{r},\tau_{p}}-\tau_{p}} \delta_{-1,X_{k}} \right)=0 \right).
\end{eqnarray*}

By the condition of the ruin takes place for the first time at time $T_{\tau_{r},\tau_{p}}$, the results of the bet at earlier times are restricted. We will show this in the following.

First, we assume that $X_{T_{\tau_{r},\tau_{p}}-\tau_{p}}=1$, then by the condition of the 
ruin at $T_{\tau_{r},\tau_{p}}$ for the first time leads to 
\[
0= U_{T_{\tau_{r},\tau_{p}}}= x+\left( \displaystyle \sum_{k=1}^{T_{\tau_{r},\tau_{p}}-\tau_{r}} X_{k} \right)-\left( \displaystyle \sum_{k=T_{\tau_{r},\tau_{p}}-\tau_{r}+1}^{T_{\tau_{r},\tau_{p}}-\tau_{p}} \delta_{-1,X_{k}} \right).
\]
Hence, the same condition requires that 
\begin{equation}
X_{T_{\tau_{r},\tau_{p}}-\tau_{r}}=-1\nonumber
\end{equation}
and
\begin{equation}
x+\left( \displaystyle \sum_{k=1}^{T_{\tau_{r},\tau_{p}}-\tau_{r}} X_{k} \right)-\left( \displaystyle \sum_{k=T_{\tau_{r},\tau_{p}}-\tau_{r}+1}^{T_{\tau_{r},\tau_{p}}-\tau_{p}-1} \delta_{-1,X_{k}} \right)=0\nonumber
\end{equation}
are simultaneously satisfied.

However, these assumptions and restriction leads to 
\begin{eqnarray*}
0&=&U_{T_{\tau_{r},\tau_{p}}}\\
&=&x+\left( \displaystyle \sum_{k=1}^{T_{\tau_{r},\tau_{p}}-\tau_{r}} X_{k} \right) -\left( \displaystyle \sum_{k=T_{\tau_{r},\tau_{p}}-\tau_{r}+1}^{T_{\tau_{r},\tau_{p}}-\tau_{p}} \delta_{-1,X_{k}} \right) \\
&\ &  (by {\ }X_{T_{\tau_{r},\tau_{p}}-\tau_{p}}=1) \\
&=&x+\left( \displaystyle \sum_{k=1}^{T_{\tau_{r},\tau_{p}}-\tau_{r}} X_{k} \right) -\left( \displaystyle \sum_{k=T_{\tau_{r},\tau_{p}}-\tau_{r}+1}^{T_{\tau_{r},\tau_{p}}-\tau_{p}-1} \delta_{-1,X_{k}} \right)\\
&=& x+ \left( \displaystyle \sum_{k=1}^{T_{\tau_{r},\tau_{p}}-\tau_{r}-1} X_{k} \right)+X_{T_{\tau_{r},\tau_{p}}-\tau_{r}}- \left( \displaystyle \sum_{k=T_{\tau_{r},\tau_{p}}-\tau_{r}+1}^{T_{\tau_{r},\tau_{p}}-\tau_{p}-1} \delta_{-1,X_{k}} \right)\\
&\ &  (by {\ }X_{T_{\tau_{r},\tau_{p}}-\tau_{r}}=-1)\\
&=& x+\left( \displaystyle \sum_{k=1}^{T_{\tau_{r},\tau_{p}}-\tau_{r}-1} X_{k} \right) - \left( \displaystyle \sum_{k=T_{\tau_{r},\tau_{p}}-\tau_{r}}^{T_{\tau_{r},\tau_{p}}-\tau_{p}-1} \delta_{-1,X_{k}} \right) \\
&=&U_{T_{\tau_{r},\tau_{p}}-1}.
\end{eqnarray*}

This means that the ruin takes place at $T_{\tau_{r},\tau_{p}}-1$, one time step earlier, which contradicts that the gambler's ruin occurs at $T_{\tau_{r},\tau_{p}}$ for the first time.

Thus, the assumption cannot be accepted and $X_{T_{\tau_{r},\tau_{p}}-\tau_{p}}\not= 1$, and must be
 $X_{T_{\tau_{r},\tau_{p}}-\tau_{p}}=-1$

Even with $X_{T_{\tau_{r},\tau_{p}}-\tau_{p}}=-1$, it is apparent that $X_{T_{\tau_{r},\tau_{p}}-\tau_{r}}=-1$ is also required. In other words, for the ruin to takes place at $T_{\tau_{r},\tau_{p}}$ for the first time, the amount of points the gambler received at that time must be negative.

Incorporating these factors, we can further reduce the expression for the ruin probability for $\theta>0$ to the following
\begin{eqnarray*}
P_{A}^{\tau_{r},\tau_{p}}(x) &=& P(U_{T_{\tau_{r},\tau_{p}}}=0)\\
&=&  P\left( x+\left( \displaystyle \sum_{k=1}^{T_{\tau_{r},\tau_{p}}-\tau_{r}} X_{k} \right)-\left( \displaystyle \sum_{k=T_{\tau_{r},\tau_{p}}-\tau_{r}+1}^{T_{\tau_{r},\tau_{p}}-\tau_{p}} \delta_{-1,X_{k}} \right)=0 \right)\\
&\ &  (by {\ }X_{T_{\tau_{r},\tau_{p}}-\tau_{p}}=-1)\\
&=&  P\left( x+\left( \displaystyle \sum_{k=1}^{T_{\tau_{r},\tau_{p}}-\tau_{r}} X_{k} \right)-\left( \displaystyle \sum_{k=T_{\tau_{r},\tau_{p}}-\tau_{r}+1}^{T_{\tau_{r},\tau_{p}}-\tau_{p}-1} \delta_{-1,X_{k}} \right) -1=0 \right)\\
&=&  P\left( (x-1)+\left( \displaystyle \sum_{k=1}^{T_{\tau_{r},\tau_{p}}-\tau_{r}} X_{k} \right)-\left( \displaystyle \sum_{k=T_{\tau_{r},\tau_{p}}-\tau_{r}+1}^{T_{\tau_{r},\tau_{p}}-\tau_{p}-1} \delta_{-1,X_{k}} \right) =0 \right)\\
\end{eqnarray*}

At this point, the initial amount of the asset is shifted by one point, but further simplification is hindered by the term containing the Kronecker's delta.
Even though the exact evaluation is not simple, we now employ an approximate assumption to replace this term by its expectation value. Namely, we assume

\[
\displaystyle \sum_{k=T_{\tau_{r},\tau_{p}}-\tau_{r}+1}^{T_{\tau_{r},\tau_{p}}-\tau_{p}-1} \delta_{-1,X_{k}} 
\approx E\left[ \displaystyle \sum_{k=T_{\tau_{r},\tau_{p}}-\tau_{r}+1}^{T_{\tau_{r},\tau_{p}}-\tau_{p}-1} \delta_{-1,X_{k}} \right]=q(\theta-1).
\]

With this assumption, the probability of ruin is approximated as
\begin{eqnarray}
P_{A}^{\tau_{r},\tau_{p}}(x) & \approx & P\left( (x-1)+\left( \displaystyle \sum_{k=1}^{T_{\tau_{r},\tau_{p}}-\tau_{r}} X_{k} \right)-q(\theta-1) =0 \right)\nonumber\\
&=& P\left( \{x-1-q(\theta-1) \}+\left( \displaystyle \sum_{k=1}^{T_{\tau_{r},\tau_{p}}-\tau_{r}} X_{k} \right)=0 \right)\nonumber
\end{eqnarray}

This equation is in the same form as Eq.(\ref{sdelay}) by identifying the initial asset is decreased by $1+q(\theta-1)$. 

Thus, this approximation leads to the ruin probability as
\begin{equation}
P_{A}^{\tau_{r},\tau_{p}}(x) \approx 
\begin{cases}
{{A - \{ x-(1+q(\theta-1))\} }\over A} & (p=q={1 \over 2})\\
\frac{(\frac{q}{p})^A - (\frac{q}{p})^{(x-(1+q(\theta-1))}}{(\frac{q}{p})^A-1} & (p \not= q)
\end{cases}
\label{tpos}
\end{equation}

This is a natural result considering that the receipt of points are more delayed than the payments. 
Hence, the gambler's asset tends to be lower at any time points, leading to a higher probability of ruin compared to the case of no delays or the same delays. 
This approximation accounts for these effects by shifting the initial assets to lower points. 
We will compare this approximation with computer simulations in the next section.

\subsection{ $\theta < 0 \quad (\tau_{r} < \tau_{p})$}

We now consider the opposite case with $\tau_{r} < \tau_{p}$.
By essentially the same arguments, we arrive at the ruin probability for this case
as follows.
\begin{eqnarray*}
P_{A}^{\tau_{r},\tau_{p}}(x) &=& P(U_{T_{\tau_{r},\tau_{p}}}=0)\\
&=& P\left( x+\left( \displaystyle \sum_{k=1}^{T_{\tau_{r},\tau_{p}}-\tau_{p}} X_{k} \right)+\left( \displaystyle \sum_{k=T_{\tau_{r},\tau_{p}}-\tau_{p}+1}^{T_{\tau_{r},\tau_{p}}-\tau_{r}-1} \delta_{1,X_{k}} \right)=0 \right)
\end{eqnarray*}

We again use the average of the term containing the Kronecker's delta.

\[
E \left [  \displaystyle \sum_{k=T_{\tau_{r},\tau_{p}}-\tau_{p}+1}^{T_{\tau_{r},\tau_{p}}-\tau_{r}-1} \delta_{1,X_{k}}  \right ]=p(-\theta -1)
\]

This leads to the following approximation for the ruin probability.
\begin{equation}
P_{A}^{\tau_{r},\tau_{p}}(x) \approx P\left( \{ x+p(-\theta -1)\}+\left( \displaystyle \sum_{k=1}^{T_{\tau_{r},\tau_{p}}-\tau_{p}} X_{k} \right)=0 \right)\nonumber
\end{equation}

Again, this equation is in the same form as Eq.(\ref{sdelay}) by identifying the initial asset is increased by $p(-\theta-1)$. 

Thus, this approximation leads to the ruin probability as
\begin{equation}
P_{A}^{\tau_{r},\tau_{p}}(x) \approx 
\begin{cases}
{{A - \{ x+p(-\theta-1)\} }\over A} & (p=q={1 \over 2})\\
\frac{(\frac{q}{p})^A - (\frac{q}{p})^{(x+p(-\theta-1))}}{(\frac{q}{p})^A-1} & (p \not= q)
\end{cases}
\label{tneg}
\end{equation}

As in the case for $\theta >0$, we can see that this is an effective approximation to account for the decrease of the ruin probability using an increase of the initial asset points.

\section{Comparison Against Computer Simulations}

In this section, we will compare our approximate results for $P_{A}^{\tau_{r},\tau_{p}}(x)$ with computer simulations. 

We will fix the following parameters:
\begin{itemize}
\item
$A = 100$
\item
$p = 9/19,\quad (q = 11/19)$
\end{itemize}

Also, we take  $10,000$ trials to obtain average values from computer simulations.

\subsection{$\theta > 0, \quad (\tau_{r} > \tau_{p})$}

For the simplicity, we set $\tau_{p} = 0$ and vary $\tau_{r}$, and initial asset points $x$. 
The results are given in the following five tables. The Column A, B are, respectively, the estimations from computer simulations, and from our approximation Eq. (\ref{tpos}).
Though data are limited due to constraints on computational times, for the ranges of
$\tau_{r} (= \theta)$, the discrepancy is less than 2 point percentiles.
\begin{table}[htbp]
\begin{center}
\begin{tabular}{|l|c|c|} \hline
$\tau_{r}$ & A &B \\  \hline
1 & 0.4687 & 0.4686 \\ 
2 & 0.4972 & 0.4972 \\
3 & 0.5354 & 0.5244 \\
4 & 0.5610 & 0.5500 \\
5 & 0.5894 & 0.5743 \\ \hline
\end{tabular}
\end{center}
\caption{The case with $x=95$}
\end{table}
\vspace{2em}

\begin{table}[htbp]
\begin{center}
\begin{tabular}{|l|c|c|} \hline
$\tau_{r}$ & A &B \\  \hline
1 & 0.6883 & 0.6862 \\ 
2 & 0.7039 & 0.7031 \\
3 & 0.7283 & 0.7192 \\
4 & 0.7422 & 0.7343 \\
5 & 0.7587 & 0.7486 \\
6 & 0.7696 & 0.7622 \\
7 & 0.7883 & 0.7750 \\
8 & 0.8036 & 0.7872 \\
9 & 0.8082 & 0.7986 \\
10 & 0.8223 & 0.8095 \\ \hline
\end{tabular}
\end{center}
\caption{The case with $x=90$}
\end{table}
\vspace{2em}

\begin{table}[htbp]
\begin{center}
\begin{tabular}{|l|c|c|} \hline
$\tau_{r}$ & A &B \\  \hline
1 & 0.8160 & 0.8147 \\ 
2 & 0.8217 & 0.8247 \\
3 & 0.8388 & 0.8341 \\
4 & 0.8490 & 0.8431 \\
5 & 0.8565 & 0.8515 \\
6 & 0.8596 & 0.8595 \\
7 & 0.8729 & 0.8671 \\
8 & 0.8779 & 0.8743 \\
9 & 0.8888 & 0.8811 \\
10 & 0.8938 & 0.8875 \\ \hline
\end{tabular}
\end{center}
\caption{The case with $x=85$}
\end{table}
\vspace{2em}

\begin{table}[htbp]
\begin{center}
\begin{tabular}{|l|c|c|} \hline
$\tau_{r}$ & A &B \\  \hline
1 & 0.9602 & 0.9619 \\ 
2 & 0.9636 & 0.9639 \\
3 & 0.9660 & 0.9659 \\
4 & 0.9682 & 0.9677 \\
5 & 0.9705 & 0.9695 \\  \hline
\end{tabular}
\end{center}
\caption{The case with $x=70$}
\end{table}
\vspace{2em}

\begin{table}[htbp]
\begin{center}
\begin{tabular}{|l|c|c|} \hline
$\tau_{r}$ & A &B \\  \hline
1 & 0.9954 & 0.9954 \\ 
2 & 0.9960 & 0.9956 \\
3 & 0.9959 & 0.9959 \\
4 & 0.9963 & 0.9961 \\
5 & 0.9965 & 0.9963 \\  \hline
\end{tabular}
\end{center}
\caption{The case with $x=50$}
\end{table}

\subsection{$\theta < 0, \quad (\tau_{r} < \tau_{p})$}

Again, for the simplicity, we set $\tau_{r} = 0$ and vary $\tau_{p}$ and initial asset points $x$. 
The results are given in the following five tables. The Column A, B are, respectively, the estimations from computer simulations, and from our approximation Eq. (\ref{tneg}).
\begin{table}[h]
\begin{center}
\begin{tabular}{|l|c|c|} \hline
$\tau_{p}$ & A &B \\  \hline
1 & 0.6461 & 0.6513 \\ 
2 & 0.6360 & 0.6335 \\
3 & 0.6241 & 0.6147 \\
4 & 0.6092 & 0.5950 \\
5 & 0.5889 & 0.5743 \\
6 & 0.5774 & 0.5525 \\
7 & 0.5571 & 0.5296 \\
8 & 0.5342 & 0.5055 \\
9 & 0.5094 & 0.4802 \\  \hline
\end{tabular}
\end{center}
\caption{The case with $x=90$}
\end{table}
\begin{table}[htbp]
\begin{center}
\begin{tabular}{|l|c|c|} \hline
$\tau_{p}$ & A &B \\  \hline
1 & 0.7950 & 0.7941 \\ 
2 & 0.7808 & 0.7836 \\
3 & 0.7728 & 0.7725 \\
4 & 0.7633 & 0.7609 \\
5 & 0.7547 & 0.7486 \\
6 & 0.7447 & 0.7358 \\
7 & 0.7321 & 0.7222 \\
8 & 0.7252 & 0.7080 \\
9 & 0.7122 & 0.6931 \\
10 & 0.7021 & 0.6774 \\ \hline
\end{tabular}
\end{center}
\caption{The case with $x=85$}
\end{table}
\begin{table}[htbp]
\begin{center}
\begin{tabular}{|l|c|c|} \hline
$\tau_{p}$ & A &B \\  \hline
1 & 0.9586 & 0.9576 \\ 
2 & 0.9528 & 0.9555 \\
3 & 0.9527 & 0.9532 \\
4 & 0.9511 & 0.9508 \\
5 & 0.9492 & 0.9483 \\ \hline
\end{tabular}
\end{center}
\caption{The case with $x=70$}
\end{table}

\section{Discussion}

We have presented an extension of gambler's ruin to include delays in gaining and losing a unit of gambler's assets. Though exact analysis of the ruin probability is difficult when there is a difference
between delays associated with gain and loss, we proposed an approximation scheme. The scheme essentially finds shifts in the initial assets to account for the effects of the delays and reduces the problem to a normal gambler's ruin with a shifted initial assets. 

These effective shifts increases(decreases) initial assets when the gaining (losing) delay is shorter leading to changes in the ruin probability. Though the comparison against computer simulations are
preliminary, our approximation can function well for small delay differences, particularly for the case that the initial asset is closer to the mid-points between two boundaries. Further analysis is left 
for the future.

\paragraph{Acknowledgments}  
This work was in part supported by research funds from Ohagi Hospital (Hashimoto, Wakayama, Japan) and NT Engineering Corporation (Takahama, Aichi, Japan).

\bigskip

\noindent\textit{Graduate School of Mathematics, Nagoya University,  Nagoya,  Japan.\\
ohira@math.nagoya-u.ac.jp}

\end{document}